\documentclass[aps,prl,notitlepage,superscriptaddress,showpacs,twocolumn]{revtex4-1}
\usepackage{graphicx,subfigure,epsfig}
 \usepackage{array,multirow}
\usepackage{dcolumn}
\usepackage{amssymb,amsmath,amsfonts,mathrsfs}
\usepackage{array}
\usepackage{times,setspace}
\usepackage{latexsym}
\usepackage{float,flafter,bm,bbm}
\usepackage{epstopdf,color,multirow}
\usepackage[colorlinks,linkcolor=blue,anchorcolor=blue,urlcolor=blue,citecolor=blue]{hyperref}
\usepackage{footnote}
\usepackage{booktabs}
\hypersetup{
    colorlinks=true,
    linkcolor=blue,
    filecolor=magenta,
    urlcolor=blue,
}

\begin{document}

\title {Asynchronous mass inversion enriched quantum anomalous Hall states in multilayer graphene}
\author{Xilin Feng}\thanks{These two authors contributed equally to this work.}\thanks{fengxilin@ust.hk}
\affiliation{Department of Physics, Hong Kong University of Science and Technology, Clear Water Bay, Hong Kong, China}
\author{Zi-Ting Sun}\thanks{These two authors contributed equally to this work.}
\affiliation{Department of Physics, Hong Kong University of Science and Technology, Clear Water Bay, Hong Kong, China}
\author{K. T. Law}\thanks{phlaw@ust.hk}
\affiliation{Department of Physics, Hong Kong University of Science and Technology, Clear Water Bay, Hong Kong, China}

\date{\today}

\begin{abstract}
Recently, multilayer graphene systems have attracted significant attention due to the discovery of a variety of intriguing phases, particularly quantum anomalous Hall (QAH) states. In rhombohedral pentalayer graphene (RPG), both QAH states with Chern number \(C = -5\) and \(C = -3\) have been observed. While the \(C = -5\) QAH state is well understood, the origin of the \(C = -3\) QAH state remains unclear. In this letter, we propose that the \(C = -3\) QAH state, as well as the topological phase transition from \(C = -3\) to \(C = -5\) state in RPG, arises from an asynchronous mass inversion mechanism driven by the interplay between trigonal warping, staggered layer order, and the displacement field: Trigonal warping splits the low-energy bands of RPG into a central touching point and three satellite Dirac cones. Meanwhile, the coexistence of the staggered layer order and displacement field induces a momentum-dependent effective mass in the low-energy bands. Consequently, mass inversions at the central touching point and the satellite Dirac cones, induced by an increasing displacement field, can occur asynchronously, leading to the formation of the \(C = -3\) QAH state and the topological phase transition from QAH state with $C=-3$ to $C=-5$. Additionally, based on this mechanism, we predict the presence of a \(C=3\) QAH state in rhombohedral tetralayer graphene (RTG), which can be detected experimentally. Furthermore, this mechanism can also be applied to Bernal tetralayer graphene (BTG), explaining the origin of the observed \(C=6\) QAH state.
\end{abstract}
\maketitle
\emph{Introduction.}---In recent years, multilayer graphene has been widely studied both theoretically and experimentally, since it provides an ideal platform for exploring several intriguing phases, such as the unconventional superconducting phase~\cite{zhou2021superconductivity,ghazaryan2021unconventional,you2022kohn,zhou2022isospin,chou2022acoustic,pangburn2023superconductivity,qin2023functional,wagner2024superconductivity,li2024tunable,yang2024diverse,vinas2024phonon,fischer2024spin,geier2024chiral,chau2024origin,wang2024chiral,wang2024chern,kim2025topological,choi2025superconductivity,zhang2025fermion,han2025signatures}, the integer and fractional quantum anomalous Hall (QAH) states~\cite{nandkishore2010quantum,zhang2010spontaneous,zhang2011spontaneous,shi2020electronic,kerelsky2021moireless,park2023topological,kwan2023moir,mondal2023quantum,yu2024moir,liu2024spontaneous,dong2024theory,han2024correlated,han2024large,sha2024observation,liu2025layer,dong2024anomalous,soejima2024anomalous,dong2024stability,kudo2024quantum,perea2024quantum,xie2024integer,lu2024fractional,ju2024fractional,huang2024fractional,das2024thermal,herzog2024moire,xie2024even,huang2024self,an2024magnetic,lu2024interaction,zhou2024fractional,aronson2024displacement,kwan2024strong,wei2025edge,lu2025extended,pichler2025single,waters2025chern,bernevig2025berry}, and the ferro-valleytricity ordered state~\cite{han2023orbital,liu2024ferro,islam2025harnessing}. Among these fascinating phases, QAH states with high Chern numbers in rhombohedral multilayer graphene (RMG) are of particular interest due to their potential applications in tunable low-dissipation electronic devices~\cite{han2024correlated,han2024large,sha2024observation,liu2025layer}. The possibility of realizing QAH states in RMG was proposed over a decade ago~\cite{zhang2011spontaneous}, based on a mechanism extended from the Haldane model~\cite{haldane1988model}, where the inversion of the Dirac mass at a single valley plays a central role~\cite{mele2015winding,weng2015quantum,liu2016quantum,chang2023colloquium}. In rhombohedral $N$-layer graphene, the low-energy physics is described by a two-band Dirac model with a $|k|^{N}$ dispersion~\cite{mccann2006landau,min2008chiral,nakamura2008electric,gruneis2008tight,koshino2009trigonal,zhang2010band,jia2013gap,mccann2013electronic}, where the relevant orbital components predominantly originate from the bottom-layer $A_1$ sublattice and the top-layer $B_N$ sublattice. Within this two-band model, the presence of layer-antiferromagnetic (LAF) order~\cite{nilsson2006electron,zhang2010spontaneous,lemonik2012competing,lang2012antiferromagnetism,scherer2012instabilities,kharitonov2012antiferromagnetic,wang2013layer,yuan2013possible,liang2022gate,jiang2024spontaneous} and Ising spin-orbit coupling (SOC)~\cite{gmitra2013spin,gmitra2015graphene,gmitra2017proximity,kochan2017model,li2019twist,naimer2021twist,zollner2023twist} enables the displacement field to drive a mass inversion in the band associated with a single spin and valley. This leads to the emergence of QAH states with Chern number $C=\pm N$~\cite{zhang2011spontaneous,liu2025layer}.

\begin{figure}[t]
    \centering
    \includegraphics[width=1\linewidth]{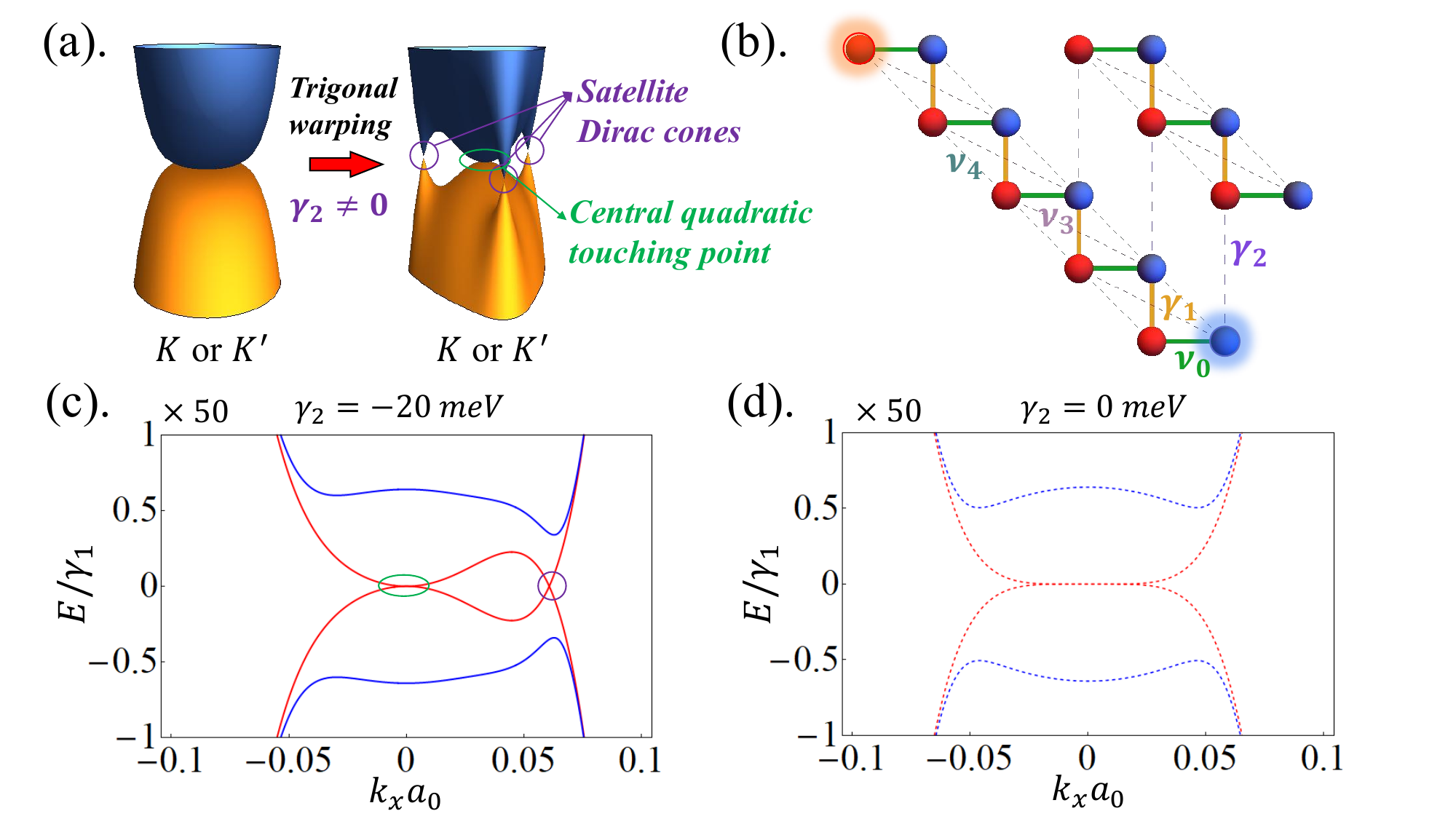}
    \caption{
    (a). The band structure for the low-energy two-band model near $K$ (or $K'$) valley without (left) and with (right) trigonal warping. The presence of trigonal warping leads to the low-energy bands of RPG split into three satellite Dirac cones and a central point with quadratic band-touching.
    (b). Side view of the rhombohedral pentalayer graphene (RPG), the blue and red atoms represent the $A$ and $B$ sublattices, respectively. The hopping parameters are labeled by different colors ($v_{i}=\sqrt{3}\gamma_{i}/2$): $\gamma_{0}=3160~meV$, $\gamma_{1}=390~meV$, $\gamma_2=-20~meV$, $\gamma_{3}=315~meV$, and $\gamma_{4}=44~meV$~\cite{koshino2009trigonal}.
    (c). The band structure of the low-energy two-band model in Eq.~(\ref{HRPG}) along $k_{y}=0$ line at $K$ valley. The red solid lines represent the bands for the case where $m=0$, $\Delta=0$, and $\gamma_{2}=-20~ meV$, while the blue solid lines represent the bands in the case where $m=-10 ~meV$, $\Delta=5~meV$ and $\gamma_{2}=-20 ~meV$. (d). The band structure of the low-energy two-band model in Eq.~(\ref{HRPG}) along $k_{y}=0$ line at $K$ valley when the trigonal warping effect is absent ($\gamma_{2}=0$). In (c) and (d), the same color of the bands indicates the same values of $m$ and $\Delta$.}
    \label{RPG0}
\end{figure}

Recently, the experimental realization of the QAH state in rhombohedral pentalayer graphene (RPG)~\cite{han2024correlated} has confirmed the predicted QAH state with Chern number $C = -5$. Surprisingly, another QAH state with $C=-3$ has also been observed in RPG, which remains unexplained by the current theoretical framework~\cite{zhang2011spontaneous,jung2011lattice,han2024correlated,liu2025layer}.
In this Letter, we identify the crucial role of the trigonal warping effect, which is usually neglected when investigating the QAH state in multilayer graphene systems~\cite{zhang2011spontaneous,jung2011lattice,han2024correlated,liu2025layer}. By incorporating trigonal warping into the low-energy effective Hamiltonian of RPG, we propose that the $C = -3$ QAH state observed in RPG arises from an asynchronous mass inversion mechanism, described as follows: In the presence of trigonal warping, the low-energy bands near the $K$ (or $K'$) valley split into three satellite Dirac cones with $C_3$ symmetry, along with a central quadratic band touching point~\cite{koshino2009trigonal}, as illustrated in Fig.\ref{RPG0}(a). When a staggered layer order~\cite{grushina2015insulating} (which is analogous to the layer antiferromagnetic (LAF) order) coexists with the displacement field, their combined effect induces a momentum-dependent effective mass in the low-energy bands of RPG. As a result, under a fixed staggered layer order, the central band-touching point and the satellite Dirac cones acquire different effective masses under the same displacement field, owing to their distinct momenta. Consequently, as the displacement field increases, mass inversions at the central and satellite band-touching points occur asynchronously. When mass inversion occurs at the three satellite Dirac cones, because each of them carries a Berry phase of \(\zeta \pi\) (with \(\zeta = \pm 1\) for the \(K\) and \(K'\) valleys, respectively), the Chern number in RPG can change by \(\pm 3\). This mechanism leads to the emergence of a QAH state with \(C = -3\) in RPG. 

Additionally, the asynchronous mass inversion mechanism also accounts for the topological phase transition from the $C=-3$ to $C=-5$ QAH state observed in RPG as the displacement field increases~\cite{han2024correlated}. After the $C=-3$ QAH state appears, a further increase in the displacement field leads to a mass inversion at the central band-touching point. Owing to its quadratic nature, the band near this point carries a Berry phase of \(2 \zeta \pi\). As a result, its mass inversion changes the Chern number by \(\pm 2\), driving a topological phase transition from the \(C = -3\) QAH state to \(C = -5\). Importantly, the asynchronous mass inversion mechanism is generally applicable to multilayer graphene systems, regardless of stacking type (rhombohedral or Bernal) or layer number. For example, our theory also explains the observed \(C=6\) QAH state~\cite{cho2025tunable} in Bernal-stacked tetralayer graphene (BTG). Our theory addresses the long-standing puzzle regarding the origin of the \(C = -3\) QAH state in RPG, and offers a possible mechanism for understanding \(C \neq \pm N, \pm 2N\) QAH states in \(N\)-layer graphene, thereby guiding future experimental explorations.

\emph{The effective model for RPG}---
The Hamiltonian for RPG without the displacement field and orders in the basis $\psi=(A_1,B_1,A_2,B_2,\dots,A_5,B_5)$ can be expressed as:
\begin{equation}\label{fullH5}
    H_{RPG}=\begin{pmatrix}
       H_0 & V & W & &   \\
        V^\dagger & H_0 & V & W &   \\
          W^\dagger & V^\dagger & H_0 & V & W   \\
             & W^\dagger & V^\dagger & H_0 & V   \\
               & & W^\dagger & V^\dagger & H_0  
        \end{pmatrix},
\end{equation}
where $A_i$ and $B_i$ represent the $A$ and $B$ sites at layer $i$. Here, $H_0$ is a $2 \times 2$ matrix that represents the intralayer hopping, while $V$ and $W$ encode the nearest and next-nearest interlayer hopping terms, respectively. Their specific form can be found in the Supplemental Material~\cite{supp}, and the hopping terms are shown in Fig.~\ref{RPG0}(b).
The interlayer potential induced by the displacement field is written as~\cite{ghazaryan2023multilayer}:
\begin{equation}\label{disD}
    H_{\Delta}=\Delta ~ Diag(D_{1},D_{2},...,D_{5}),
\end{equation}
where \( Diag (...) \) denotes a diagonal matrix whose diagonal elements are listed in the brackets, $D_{l}=(1-\frac{(l-1)}{2})\sigma_{0}$ (for $l=1,2,...,5$), $\sigma_{0}$ is the $2 \times 2$ identity matrix, \( l \) is the layer index, and $\Delta$ is the amplitude of this interlayer potential.
Additionally, the staggered interlayer order is:
\begin{equation}\label{interorder1}
\begin{aligned}
    H_{m}=m ~Diag(U_{1},U_{2},O,-U_{4},-U_{5}),        
\end{aligned}
\end{equation}
where $U_{l}=(-1)^{l-1}\sigma_{0}$, $O$ is the $2 \times 2$ zero matrix, and \( m \) is the amplitude of this order. To capture the key physical mechanism underlying the QAH states in RPG, we derive a low-energy two-band model from the full Hamiltonian ($H^{full}=H_{RPG}+H_{\Delta}+H_{m}$), which incorporates the displacement field and staggered layer order. Under the basis $(A_{1}, B_{5})$, the low-energy model is expressed as:
\begin{equation}\label{HRPG}
     H=\begin{pmatrix} M_{5,eff}(k)& X_{5}(\mathbf{k})\\ X_{5}^{\dagger}(\mathbf{k}) & -M_{5,eff}(k) \end{pmatrix},
\end{equation}
where $X_{5}(\mathbf{k})$ is the effective hopping between $A_{1}$ and $B_{5}$. The specific form of $X_{5}(\mathbf{k})$ is written as:
\begin{equation}\label{hopping5}
    X_{5}(\mathbf{k})= \frac{v_0^{5}}{\gamma_{1}^{4}}(a_{0}\xi_{\mathbf{k}}^{\dagger})^5+\frac{3v_{0}^2}{\gamma_{1}^2}(a_{0}\xi_{\mathbf{k}}^{\dagger})^2(\frac{\gamma_{2}}{2}).
\end{equation}
In Eq.~(\ref{hopping5}), $\xi_{\mathbf{k}}=\zeta k_{x}+i k_{y}$~($\zeta=\pm$ for $K$ and $K'$ valley), $a_{0}$ is the lattice constant of graphene, $v_{0}$ represents the intralayer hopping, $\gamma_{1}$ represents the hopping between dimerized atoms belonging to adjacent layers, and $\gamma_{2}$ represents the next-nearest interlayer hopping which induces the trigonal warping effect. The band structures of the low-energy effective model with ($\gamma_{2} \neq 0$) and without ($\gamma_{2}=0$) the trigonal warping effect along $k_{y}=0$ line are shown in Fig~\ref{RPG0}~(c) and (d), respectively. 

Additionally, the effective mass $M_{5,\text{eff}}(k)$ of the two-band model for RPG is given by:
\begin{equation}\label{effm5-1}
    M_{5,\text{eff}}(k)= (m+\Delta)-\frac{(2 m-\Delta)v_{0}^2a_{0}^2k^2}{2\gamma_{1}^2},
\end{equation}
where $m$ is the amplitude of the staggered layer order, $\Delta$ is the interlayer potential induced by the displacement field, and $k$ is the amplitude of the momentum measured from the $K$ (or $K'$) point of the Brillouin zone (BZ). As we will show below, the presence of the trigonal warping term \(\gamma_2\) and the momentum-dependent effective mass account for the asynchronous mass inversion mechanism.

\begin{figure}[t]
    \centering
    \includegraphics[width=1\linewidth]{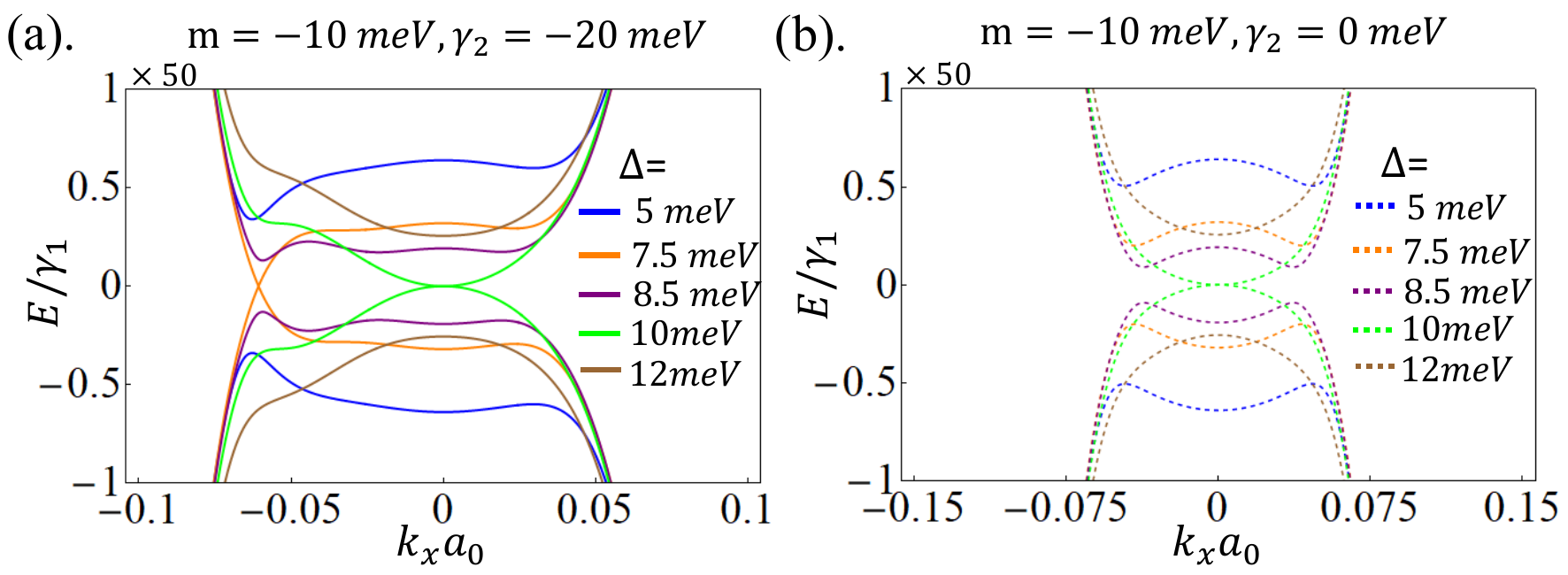}
    \caption{(a). The evolution of the band structure for the low-energy two-band model Eq.~(\ref{HRPG}) along $k_{y}=0$ line with increasing interlayer potential at $K'$ valley, in the presence of staggered layer order Eq.~(\ref{interorder1}), is shown, where the parameters are $v_{0}=\frac{\sqrt{3}}{2} \times 3160~ meV $, $\gamma_{1}=390~meV$, $m=m_{2}=-10 ~meV$ and $\gamma_{2}=-20 ~meV$. (b). The evolution of the band structure with increasing interlayer potential in the case where $\gamma_{2}=0$ (without trigonal warping). All other parameters are the same as in (a).}
    \label{evolution}
\end{figure}

\emph{Quantum anomalous Hall states in RPG}---The effective mass given in Eq.~(\ref{effm5-1}) is momentum-dependent. At the central point of \(K\) or \(K'\) valley, the effective mass is expressed as \(m + \Delta\). It changes sign when the interlayer potential satisfies \(\Delta = \Delta_{1} = -m\) (assuming \(m<0\)). At the point with $k \neq 0$, the effective mass changes sign at $\Delta = \Delta_{2}= \frac{2\epsilon - 1}{1 + \epsilon}m$, where $\epsilon = \frac{v_0^2 a_0^2 k^2}{2\gamma_1^2}$. When \(\epsilon\) satisfies the condition \(\epsilon < \frac{1}{2}\), it follows that \(0 < \Delta_2 < \Delta_1\). Therefore, as the displacement field increases, the effective masses of the three satellite Dirac cones change sign first at \(\Delta = \Delta_2\), followed by the sign change of the effective mass at the central quadratic touching point at \(\Delta = \Delta_1\). The evolution of the band structure with increasing interlayer potential ($\Delta$) for $K'$ valley is shown in Fig.~\ref{evolution}~(a). 
To explain the appearance of the QAH state with $C=-3$, we consider staggered layer orders with amplitudes \(m_1\), \(-m_1\), \(m_2\), and \(-m_2\) for the four flavors \(K\uparrow\), \(K\downarrow\), \(K'\uparrow\), and \(K'\downarrow\), respectively, where \(m_1 < m_2 < 0\). When the interlayer potential satisfies \(\Delta = \Delta_2 =\frac{2\epsilon - 1}{1 + \epsilon}m_{2}\), a mass inversion occurs at the satellite Dirac cones, causing the Chern number of the valence band with spin-up in the \(K'\) valley (flavor \(K'\uparrow\)) to change from \(\frac{5}{2}\) to \(-\frac{1}{2}\). Meanwhile, the Chern numbers for the valence band with the other three flavors—\(K\uparrow\), \(K\downarrow\), and \(K'\downarrow\)—are \(-\frac{5}{2}\), \(\frac{5}{2}\), and \(-\frac{5}{2}\), respectively, and remain unchanged.
Thus, the QAH state with \(C = -3\) appears. When the interlayer potential further increases beyond \(\Delta_1 = -m_{2}\), the effective mass at the central band-touching point for the \(K'\uparrow\) flavor changes sign, causing the Chern number of its valence band to become $-\frac{5}{2}$. As a result, a QAH state with \(C = -5\) appears.

For comparison, we also present the evolution of the band structure with increasing interlayer potential in the case where $\gamma_2 = 0$, as shown in Fig.~\ref{evolution}(b). As the interlayer potential increases, the effective mass at the central point of the $K'$ valley changes sign at $\Delta = \Delta_1 = -m$, without any additional gap closing. Furthermore, since the band-touching at this point follows a $k^5$ dispersion, it carries a Berry phase of $5\pi$. Therefore, in the absence of trigonal warping, increasing the interlayer potential induces a single topological phase transition from the $C=0$ state to a QAH state with $C=-5$.
In addition, if there is no staggered layer order ($m=0$), although this effective mass (Eq.~\ref{effm5-1}) remains momentum-dependent, its sign is only determined by the sign of the interlayer potential $\Delta$ at both $k=0$ and $k\neq 0$ points. Therefore, increasing the interlayer potential (from negative to positive) will simultaneously induce mass inversion at both the central point and the satellite Dirac cones.

\begin{figure}[b]
    \centering
    \includegraphics[width=1\linewidth]{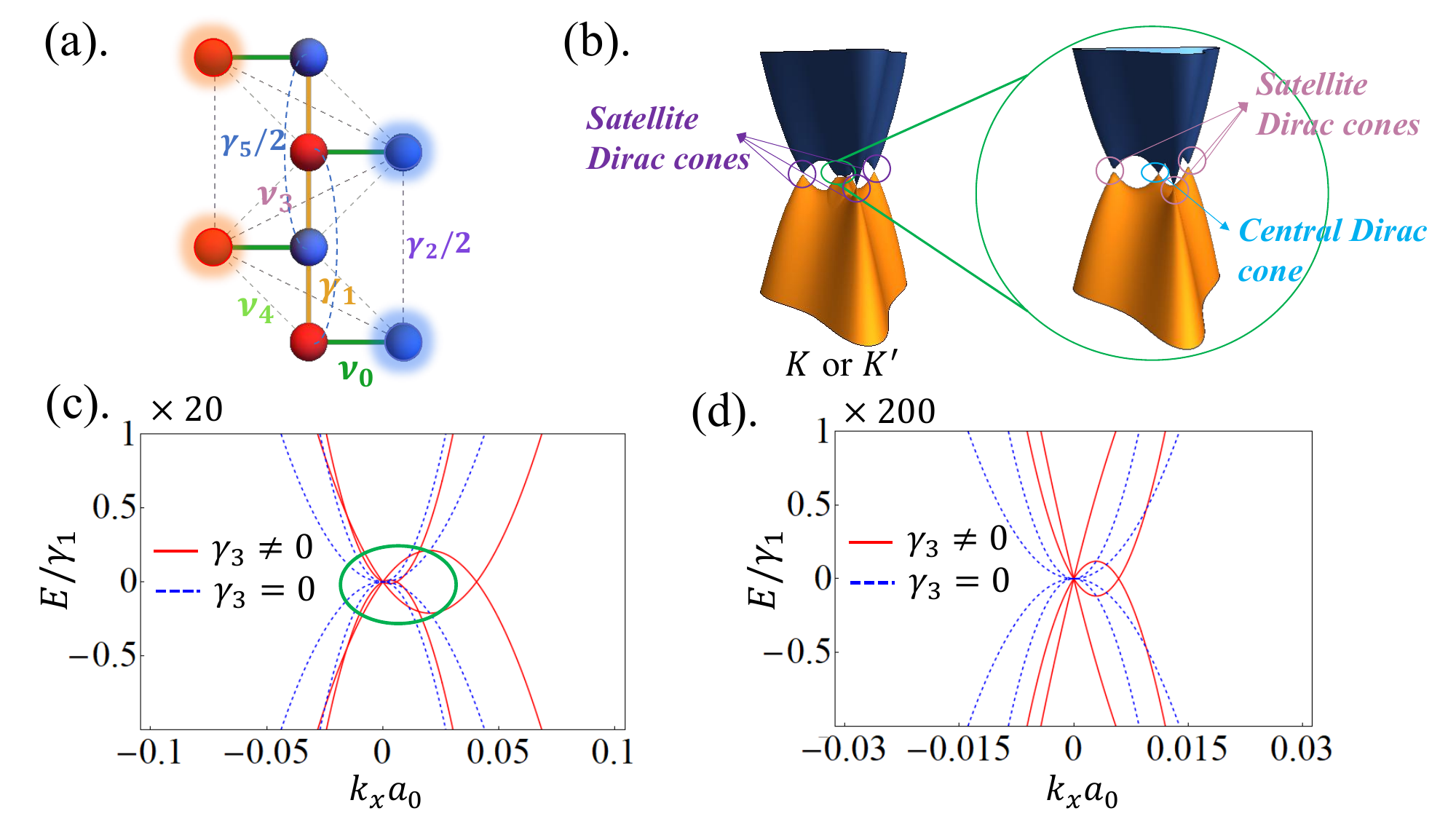}
    \caption{(a). Side view of the Bernal tetralayer graphene (BTG), the blue and red atoms represent the $A$ and $B$ sublattices, respectively. The hopping parameters are indicated by different colors ($v_{i}=\frac{\sqrt{3}}{2}\gamma_{i}$): $\gamma_{0}= 3000~ meV$, $\gamma_{1}=400~ meV$, $\gamma_{2}=-20~meV$, $\gamma_{3}=300 ~meV$, and $\gamma_{4}=\gamma_{5}=40~meV$~\cite{koshino2011landau}. Here, the displacement field and staggered layer orders are set to zero: $\Delta=0$, and $m=0$. (b). The band structure near $K$ (or $K'$) valley in the presence of the trigonal warping. The bands in the green circle show the enlarged view near $K$ (or $K'$) point. (c). The band structure for the low-energy effective model along $k_{y}=0$ line with and without $\gamma_{3}$, which are shown by red solid and blue dashed lines, respectively. Here, we set $\Delta=0$ and $m=0$. (d). Enlarged view of the band structure for the region marked by the green circle in (c).}
    \label{fig_BTG}
\end{figure}

Furthermore, to make the analysis more practical, we investigate the band structure and topological phase transition using the full Hamiltonian of RPG (Eq.~\ref{fullH5}), taking into account the staggered layer order and displacement field. Although the additional trigonal warping term, \(\gamma_3\), causes a slight further splitting at the central band touching point in the \(K\) (or \(K'\)) valley, and the interlayer hopping term, \(\gamma_4\), introduces particle-hole asymmetry, the asynchronous mass inversion mechanism and the phase diagram for RPG remain unaffected. The results of the full model calculations are shown in Supplemental Material~\cite{supp}.

\emph{Generalizing to other rhombohedral multilayer graphene systems}---To verify the universality of the asynchronous mass inversion mechanism in multilayer graphene systems and pave the way for experimental verification, we derive the low-energy effective model for rhombohedral \(N\)-layer graphene (\(N \geq 3\)) in the presence of staggered layer order and displacement field (see Supplemental Material~\cite{supp}). By applying this model to rhombohedral tetralayer graphene (RTG), we can get the effective mass for RTG (\(m<0\)):
\begin{equation}
    M_{4,\text{eff}}(k)=(m+\Delta)-\frac{(3m-\Delta)v^2a_{0}^2k^{2}}{3\gamma_{1}^{2}}.
\end{equation}
Thus, the mass-inversion conditions for the central Dirac cone and the satellite Dirac cones are \(\Delta = \Delta_1 = -m\) and \(\Delta = \Delta_2 = \frac{3\epsilon' - 1}{1 + \epsilon'} m\), respectively. Under the condition \(0 < \epsilon' = \frac{v_0^2 a_0^2 k^2}{3 \gamma_1^2} < \frac{1}{3}\), we derive that \(\Delta_2 < \Delta_1\), which predicts the emergence of a \(C = 3\) QAH state prior to the experimentally observed \(C = 4\) QAH state as the displacement field increases in RTG.

\begin{figure}[b]
\hspace*{-0.4cm}
        \centering
        \includegraphics[width=0.52\textwidth]{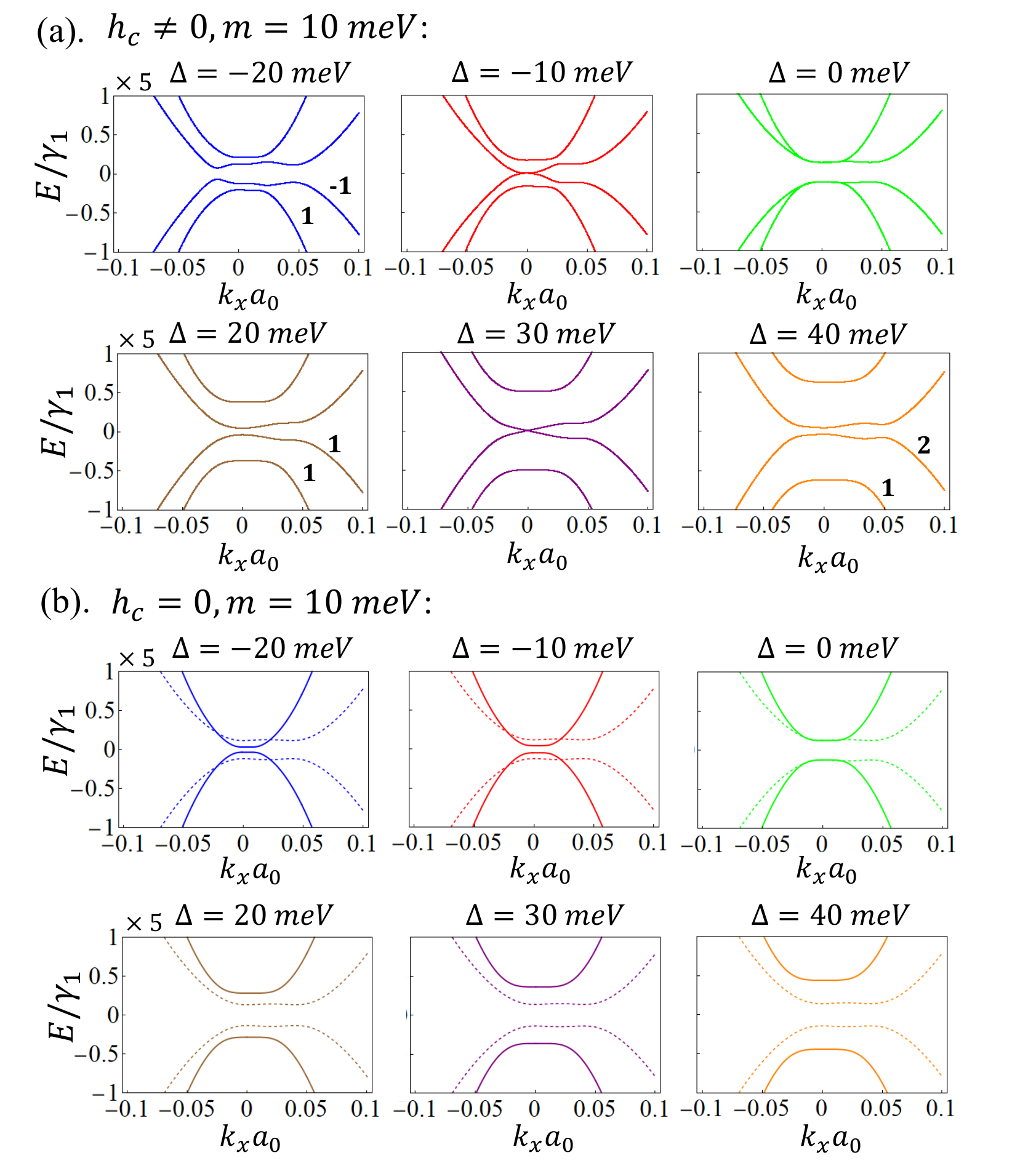}
        \caption{(a). The evolution of the band structure for the low-energy effective model of Bernal tetralayer graphene (BTG) with the increasing displacement field in $K$ valley. The Chern number for each valence band is indicated near the respective band. The parameters are $v_{0}=\frac{\sqrt{3}}{2}\times 3000~ meV$, $\gamma_{1}=400~ meV$, $v_{3}=\frac{\sqrt{3}}{2}\times 300 ~meV$, and $m=10~meV$. (b). The evolution of the band structure for the low-energy effective model of BTG with increasing displacement field, in the absence of inter-bilayer coupling ($h_{c}=0$). The solid and dashed lines represent the bands of the first and the second bilayer graphenes, respectively.}
        \label{btgband}
\end{figure}

\emph{The effective model for Bernal tetralayer graphene}---Next, we apply the asynchronous mass inversion mechanism to explain the recently observed $C=6$ QAH state in BTG~\cite{cho2025tunable}. The side view and hopping parameters for BTG are shown in Fig.~\ref{fig_BTG}~(a), and the staggered layer order is written as:
\begin{equation}\label{interorder2}
    H^{B}_{m}=m ~Diag(U_{1},U_{2},U_{3},U_{4}),
\end{equation}
where $U_{l}=(-1)^{l-1}\sigma_{0}$, with $l$ labeling the layer index. Then the low-energy effective model for BTG can be derived as~\cite{koshino2011landau}:
\begin{equation}\label{HBTG0}
    H=\begin{pmatrix}
       h_{\text{eff}}(\lambda_{1})+h_{\Delta}(s_{1,1},t_{1,1}) & h_{c}(s_{1,3},t_{1,3}) \\
        h_{c}(s_{1,3},t_{1,3}) & h_{\text{eff}}(\lambda_{3})+h_{\Delta}(s_{3,3},t_{3,3})   
        \end{pmatrix},
\end{equation}
where 
\begin{equation*}
    h_{eff}(\lambda)=\begin{pmatrix} 0&\lambda v_3 \xi_{\mathbf{k}}-\frac{1}{\lambda \gamma_{1}} v^2_0 (\xi_{\mathbf{k}}^{\dagger})^2 \\ \lambda v_3 \xi_{\mathbf{k}}^{\dagger}-\frac{1}{\lambda \gamma_{1}} v^2_0 (\xi_{\mathbf{k}})^2 & 0 \end{pmatrix},
\end{equation*}
$h_{\Delta}(s,t)=\begin{pmatrix} s\Delta+m&0 \\ 0 & t\Delta-m \end{pmatrix}$, and $h_{c}(s,t)=\begin{pmatrix} s\Delta&0 \\ 0 & t\Delta \end{pmatrix}$. The parameters in Eq.~(\ref{HBTG0}) are $\lambda_{1}=(\sqrt{5}-1)/2$, $\lambda_{3}=(\sqrt{5}+1)/2$, $s_{1,1}=-t_{1,1}=\frac{1}{15}(5+2\sqrt{5})$, $s_{3,3}=-t_{3,3}=\frac{1}{15}(5-2\sqrt{5})$, and $s_{1,3}=t_{1,3}=\frac{4}{3\sqrt{5}}$.
The band structure near $K$ (or $K'$) point for the low-energy effective model of BTG with trigonal warping is shown in Fig.~\ref{fig_BTG}~(b). In this figure, the displacement field and staggered layer order are set to zero: \( \Delta = 0 \) and \( m = 0 \).

\emph{Topological phase transitions in Bernal tetralayer graphene}---The Hamiltonian in Eq.~(\ref{HBTG0}) can be interpreted as two coupled bilayer graphenes with different hopping parameters. In the following discussion, we refer to these two bilayer graphene systems as the first ( \( H_{\text{eff}}^{1} = h_{\text{eff}}(\lambda_{1}) + h_{\Delta}(s_{1,1}, t_{1,1}) \) ) and the second ( \( H_{\text{eff}}^{2} = h_{\text{eff}}(\lambda_{3}) + h_{\Delta}(s_{3,3}, t_{3,3}) \) ) bilayer graphene, respectively. Due to the trigonal warping, the energy bands of both the first and second bilayer graphene split into one central Dirac cone and three satellite Dirac cones at the \(K\) and \(K'\) valleys. The central Dirac cone and the satellite Dirac cone possess Berry phases of \(-\zeta \pi\) and \(\zeta \pi\), respectively, where $\zeta=\pm$ corresponds to the $K$ and $K'$ valleys. Moreover, since \(\lambda_{3} > \lambda_{1}\), the momentum amplitude of the three satellite Dirac cones in the second bilayer graphene (\( k_{2} \)) is larger than that in the first bilayer graphene (\( k_{1} \)), as shown in Fig.~\ref{fig_BTG}~(b). Fig.~\ref{fig_BTG}~(c) and (d) show the band structures along the \( k_y = 0 \) line for BTG with \( \Delta = 0 \) and \( m = 0 \), explicitly demonstrating the trigonal warping effect. Additionally, as shown in the low-energy effective model (Eq.~(\ref{HBTG0})), the coupling strength between the two bilayer graphenes ($h_{c}$) is determined by the displacement field. For a fixed staggered layer order, the evolution of the band structure for BTG with the increasing displacement field is illustrated in Fig.~\ref{btgband}~(a).

We next analyze the mass inversion process as the displacement field increases in BTG. Based on the effective masses ($h_{\Delta}$) of the first and second bilayer graphenes (in Eq.~\ref{HBTG0}), we conclude that for \( m > 0 \), when the interlayer potential satisfies \( -3m < \Delta < 0 \), the amplitude of the effective mass for the first bilayer graphene, \( M_1 = |s_{1,1}\Delta + m| \), is smaller than that for the second bilayer graphene, \( M_2 = |s_{3,3}\Delta + m| \), i.e., \( M_1 < M_2 \). As a result, the energy gap between the valence and conduction bands in the first bilayer graphene is smaller than that in the second bilayer graphene. Consequently, for \( m > 0 \), when \( -3m < \Delta < 0 \), the topological phase transition arises from mass inversion in the band whose component is from the first bilayer graphene. Due to the small momentum amplitude of the three satellite Dirac cones in the first bilayer graphene, the effective masses of the central Dirac cone and the satellite Dirac cones invert simultaneously, resulting in a change of \( \pm 2 \) in the Chern number of the low-energy valence band. Conversely, for \(\Delta > 0\), since \(M_1 > M_2\), the energy gap in the second bilayer graphene becomes smaller than that in the first bilayer graphene. Therefore, when \( \Delta > 0 \), the topological phase transition arises from mass inversion in the band whose component is from the second bilayer graphene. Due to the large momentum amplitude of the satellite Dirac cones in the second bilayer graphene, the mass inversion only occurs at the central Dirac cone. This results in a change of \( \pm 1 \) in the Chern number. The band structures corresponding to two bilayer graphenes with $m>0$ are shown in Fig.~\ref{btgband}~(b). On the other hand, for the case \( m < 0 \), a similar result can be obtained: when \( \Delta < 0 \) or \( \Delta >-3m \), \( M_1 > M_2 \), the Chern number changes by $\pm 1$; while for \( 0 < \Delta < -3m \), \( M_1 < M_2 \), the Chern number changes by $\pm 2$.

\emph{Quantum anomalous Hall states in BTG}---To investigate the QAH states in BTG, we assume that the staggered orders for the four flavors \(K \uparrow\), \(K \downarrow\), \(K' \uparrow\), and \(K' \downarrow\) are given by \(H^{B}_{m}\), \(H^{B}_{m}\), \(H^{B}_{-m}\), and \(H^{B}_{-m}\), respectively, with \(m > 0\). When \(\Delta > 3m\), the total Chern number of valence bands in the \(K\) valley for each spin is 3, while in the \(K'\) valley for each spin, it is 0. This results in a QAH state in BTG with a total Chern number of \(C = 6\).  
Additionally, by using the full Hamiltonian of BTG~\cite{ghazaryan2023multilayer,chen2023gate}, we investigate its band structure and topological phase transitions. While the band structure differs from that of the low-energy effective model and the topological phase transition varies with different amplitudes of the staggered layer order \(m\), the QAH state with a total Chern number of \(C = 6\) persists (see Supplemental Material for details~\cite{supp}).

\emph{Conclusion and Discussion}--- In summary, we propose that for rhombohedral \(N\)-layer graphene, the emergence of QAH states with a Chern number \(C \neq \pm N, \pm 2N\) can be attributed to an asynchronous mass inversion mechanism, which arises from the interplay between trigonal warping, staggered layer order, and displacement field. The trigonal warping splits the low-energy bands of this multilayer graphene into the central touching point and satellite Dirac cones in the \(K\) (or \(K'\)) valley. With staggered layer order and increasing displacement fields, mass inversion can occur asynchronously at the central touching point and the satellite Dirac cones, leading to Chern number changes that are not necessarily equal to \(\pm N\). Within this mechanism, we explain the origin of the $C=-3$ QAH state in RPG~\cite{han2024correlated}. In addition, we predict the appearance of the QAH state with \(C = 3\) in RTG, which is experimentally detectable. Interestingly, the asynchronous mass inversion mechanism can also be applied to BTG, revealing the emergence of the \(C = 6\) QAH state, which has been observed experimentally~\cite{cho2025tunable}. Moreover, it is worth noting that in experiments~\cite{han2024correlated,cho2025tunable}, QAH states are observed only under the application of a small, finite magnetic field. This observation may be attributed to the stabilization of valley-polarized orders, resulting from the coupling between orbital magnetization~\cite{thonhauser2005orbital,xiao2005berry,ceresoli2006orbital,shi2007quantum,xiao2010berry,bianco2013orbital,bianco2016orbital,zhu2020voltage} and the applied magnetic field. This coupling remains to be explored in future studies.

\emph{Acknowledgements}---The authors thank Long Ju and Fan Zhang for inspiring discussions. K. T. L. acknowledges the support of the Ministry of Science and Technology, China, and Hong Kong Research Grant Council through Grants No. 2020YFA0309600, No. RFS2021-6S03, No. C6025-19G, No. AoE/P-701/20, No. 16310520, No. 16307622, and No. 16309223.

  \bibliography{ref.bib}

\begin{widetext}

\setcounter{figure}{0}  
\setcounter{section}{0}
\renewcommand\thefigure{S\Alph{section}\arabic{figure}}
\setcounter{equation}{0}
\renewcommand\theequation{S\arabic{equation}}

\newpage
\section*{\bf{\uppercase\expandafter{Supplemental Material NOTE I: rhombohedral multilayer graphene}}}

\subsection{I-A. Low-energy effective model for rhombohedral multilayer graphene ($N\geq 3$)}\label{LEM}
In this section, we derive the low-energy effective two-band model for rhombohedral multilayer graphene (RMG), incorporating the effects of trigonal warping, displacement field, and staggered layer order. The basis for the RMG is expressed as:
\begin{equation}
    \psi=(A_1,B_1,A_2,B_2,\dots,A_N,B_N),
\end{equation}
where $A_{i}$ and $B_{i}$ represent the $A$ and $B$ sublattice in the $i$ layer, and $N$ denotes the number of layers (with the assumption that $N \geq 3$). Therefore, the matrix form of the Hamiltonian for RMG is written as:
\begin{equation}\label{fullHRMG}
    H=\begin{pmatrix}
       H_0 & V & W & & & & \\
        V^\dagger & H_0 & V & W & & & \\
          W^\dagger & V^\dagger & H_0 & V & W & & \\
             & W^\dagger & V^\dagger & H_0 & V & W & \\
               & & \ddots & \ddots & \ddots & \ddots & \ddots
        \end{pmatrix},
\end{equation}
where the intralayer hopping is expressed by $H_{0}=\begin{pmatrix} 0&v_0 \xi_{\mathbf{k}}^{\dagger}\\v_0\xi_{\mathbf{k}} & 0 \end{pmatrix}$, and $\xi_{\mathbf{k}}=\zeta k_x+i k_y$ ($\zeta= \pm 1$ representing the K and K' valley, respectively). The nearest interlayer hopping is represented by $V=\begin{pmatrix} -v_4 \xi_{\mathbf{k}}^{\dagger}& v_3 \xi_{\mathbf{k}} \\ \gamma_1 & -v_4 \xi_{\mathbf{k}}^{\dagger} \end{pmatrix}$, and the next-nearest interlayer hopping is written as $W=\begin{pmatrix} 0&\gamma_2/2 \\ 0 & 0 \end{pmatrix}$. The side view of rhombohedral multilayer graphene and the hopping terms are labeled in Fig.\ref{fig_g3}~(a)~($v_{i}=\sqrt{3}\gamma_{i}/2$, in this figure, the pentalayer graphene is shown as an example). 

\begin{figure}[h]
    \centering
    \includegraphics[width=0.65\linewidth]{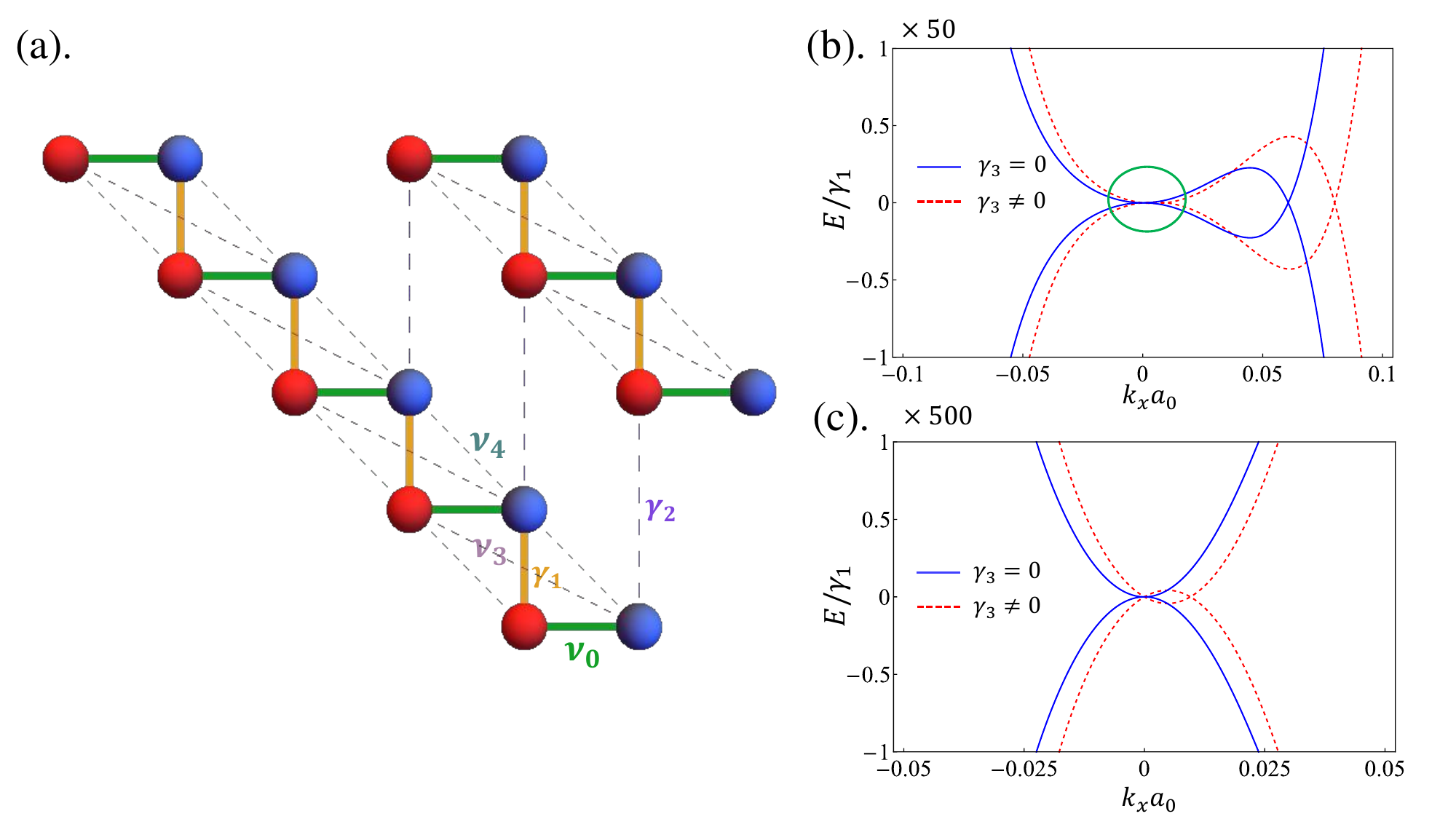}
    \caption{(a). Side view of the rhombohedral pentalayer graphene (RPG), the blue and red atoms represent the $A$ and $B$ sublattices, respectively. The hopping parameters are indicated by different colors: $\gamma_{0}=3160~ meV$, $\gamma_{1}=390~ meV$, $\gamma_{2}=-20~ meV$, and $\gamma_{4}=44 ~meV$. Here, the displacement field and staggered layer orders are set to zero: $\Delta=0$, and $m=0$. (b). The band structure for the low-energy effective model with and without $\gamma_{3}=315 ~meV$, which are shown by red dashed and blue solid lines, respectively. (c). Enlarged view of the band structure for the region marked by the green circle in (b).}
    \label{fig_g3}
\end{figure}

The Hamiltonian for staggered layer order is expressed as:
\begin{equation}\label{interorder}
    H_{m}=\begin{cases} m~Diag(U_1,...,U_{(N-1)/2},O,-U_{(N+3)/2},...,-U_{N}) & if \        \ \text{N $\in$ odd}\\
                        m~Diag(U_{1},...,U_{l},...,U_{N}) & if \        \ \text{N $\in$ even}  \end{cases},
\end{equation}
where $U_{l}=(-1)^{l-1}\begin{pmatrix}
    1&0\\0&1
\end{pmatrix}$,
$O=\begin{pmatrix}
    0&0\\0&0
\end{pmatrix}$, $l$ is the layer index, and $m$ is the amplitude of this order. \( Diag (...) \) denotes a diagonal matrix whose diagonal elements are listed in the brackets. In addition, the Hamiltonian for interlayer potential induced by the displacement field can be expressed as:
\begin{equation}\label{displacement}
    H_{\Delta}=\Delta ~ Diag(D_{1},D_{2},...,D_{N}),
\end{equation}
where $D_{l}=\begin{pmatrix}
    1-\frac{2(l-1)}{N-1}&0\\0&1-\frac{2(l-1)}{N-1}
\end{pmatrix}$ (for $l=1,2,...,N$), and $\Delta$ is the amplitude of this interlayer potential.

Then, the effective two-band Hamiltonian under the basis $(A_{1}, B_{N})$ can be obtained by performing perturbation theory. Near $K$ and $K'$ valleys, due to the relation: $\gamma_1, |v_{0} \xi_{\mathbf{k}}| >> |v_3 \xi_{\mathbf{k}}|, |v_4 \xi_{\mathbf{k}}|$, we can simplify the Hamiltonian by setting $\gamma_3=0$ and $\gamma_4=0$ in the result of the perturbation theory. Therefore, the Hamiltonian can be written as:
\begin{equation} \label{slowRMG}
    H=\begin{pmatrix} M_{N,eff}(k)& X_{N}(\mathbf{k})\\ X_{N}^{\dagger}(\mathbf{k}) & -M_{N,eff}(k) \end{pmatrix},
\end{equation}
where the effective mass term $M_{N,eff}(k)$ is momentum-dependent and can be expressed as:
\begin{equation}\label{effmass}
\begin{aligned}
        M_{N,eff}(k)=
    (m+\Delta)-\frac{((N-1) m-(N-3)\Delta)v_{0}^2a_{0}^2k^2}{(N-1)\gamma_{1}^2}.
\end{aligned}
\end{equation}
In this effective mass (Eq.~(\ref{effmass})), $m$ is the amplitude of the staggered layer order, $\Delta$ is the amplitude of the interlayer potential induced by the displacement field, $a_{0}$ is the lattice constant of the graphene, and $k^2=\xi_{\mathbf{k}}\xi^{\dagger}_{\mathbf{k}}=|k_{x}+ik_{y}|^2$. As discussed in the main text, the combination of trigonal warping and the momentum dependence of the effective mass is responsible for the appearance of additional quantum anomalous Hall (QAH) states with Chern number $C\neq \pm N, \pm 2N$ in rhombohedral $N$-layer graphene.
The effective hopping term $X_{N}(\mathbf{k})$ for the two-band model is written as~\cite{koshino2009trigonal}:
\begin{equation}\label{hopping}
    X_{N}(\mathbf{k})=\sum_{\{n_1,n_2\}}\frac{(n_1+n_2)!}{n_1!n_2!}\frac{1}{(-\gamma_1)^{n_1+n_2-1}}(v_{0}\xi_{\mathbf{k}})^{n_1}(\frac{\gamma_2}{2})^{n_2},
\end{equation}
where $n_1$ and $n_2$ are non-negative integers, and they satisfy $n_1+3 n_2=N$.

In rhombohedral pentalayer graphene (RPG), the band structure of the low-energy effective model under different displacement fields and a fixed staggered layer order is shown in Fig.2 of the main text. As the displacement field increases, the effective mass term for the three satellite Dirac cones first changes sign at \(\Delta = -m_I < -m\) (with \(m < 0\)), resulting in a topological phase transition with a change in the Chern number of $\pm 3$. As the displacement field continues to increase, the effective mass for the central band-touching point (quadratic touching) also changes sign at \(\Delta = -m\), leading to a second topological phase transition with a change in the total Chern number of $\pm 2$.

\subsection{I-B. The band structure and topological phase transition for the full Hamiltonian of RPG}
In this section, to make the analysis more practical, we calculate the band structure of RPG using the full Hamiltonian shown in Eq.~(\ref{fullHRMG}) instead of the low-energy effective model, accounting for the staggered layer order and displacement field. The evolution of the bands with increasing displacement field for a fixed staggered layer order is shown in Fig.~\ref{figs3}. Although the band structure differs from the low-energy two-band model, the asynchronous mass inversion behavior remains unchanged. Therefore, the low-energy two-band model can capture the key physical mechanism underlying the formation of the QAH states in RPG.

\begin{figure}[h]
    \centering
    \includegraphics[width=0.55\linewidth]{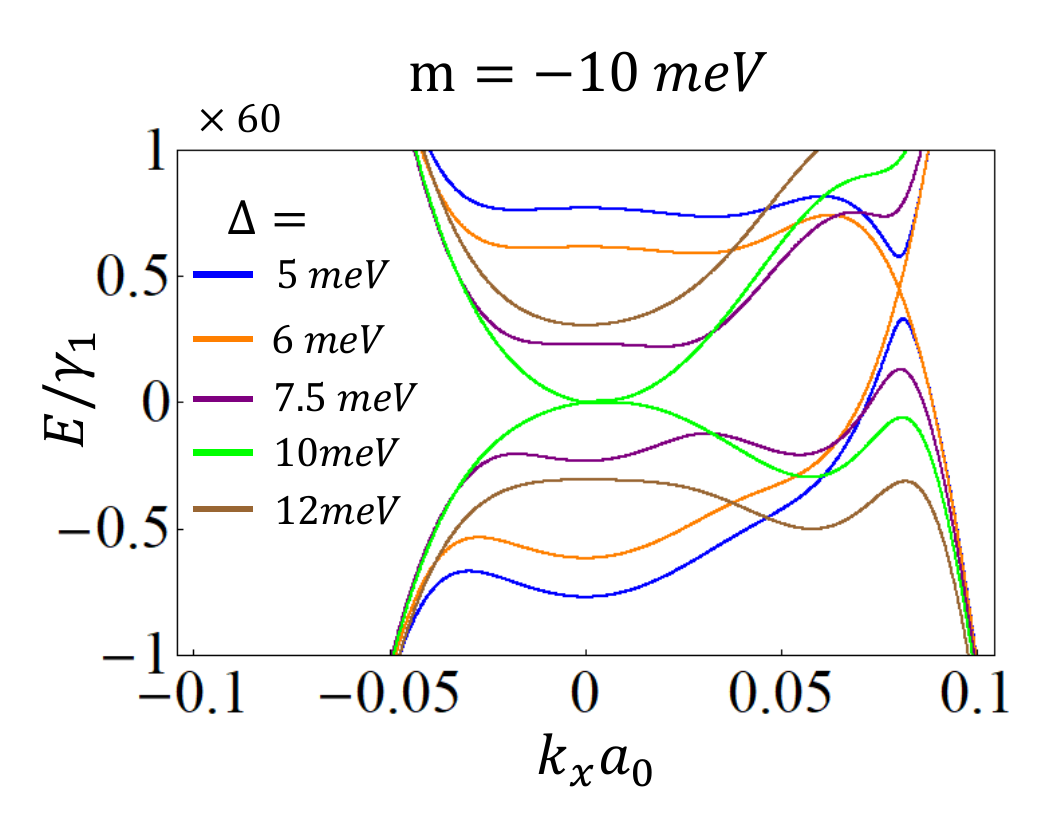}
    \caption{The evolution of the band structure for a fixed staggered layer order $m=-10 meV$ with increasing $\Delta$. Other parameters in are: $v_{0}=\frac{\sqrt{3}}{2} \times 3160~ meV $, $\gamma_{1}=390~meV$, $\gamma_{2}=-20~meV$, $v_{3}=\frac{\sqrt{3}}{2}\times 315~meV$, and $v_{4}=\frac{\sqrt{3}}{2}\times 40~meV$.}
    \label{figs3}
\end{figure}

\subsection{I-C. The effect of trigonal warping term $\gamma_3$ in rhombohedral pentalayer graphene}
In this section, we examine the effect of the trigonal warping term $\gamma_3$ in rhombohedral pentalayer graphene (RPG). 
By using the perturbation theory mentioned in Sec.I-A, the two-band effective Hamiltonian for RPG can be written as:
\begin{equation}
     H=\begin{pmatrix} M_{5,eff,2}(k)& X_{5,2}(\mathbf{k})\\ X_{5,2}^{\dagger}(\mathbf{k}) & -M_{5,eff,2}(k) \end{pmatrix},
\end{equation}
where the effective mass term is:
\begin{equation}
    M_{5,eff,2}(k)=(m+\Delta)-\frac{(2 m-\Delta)v_{0}^2k^2}{2\gamma_{1}^2},
\end{equation}
and the effective hopping $X_{5,2}(\mathbf{k})$ is:
\begin{equation}
    X_{5,2}(\mathbf{k})= \frac{v_0^{5}}{\gamma_{1}^{4}}(\xi_{\mathbf{k}}^{\dagger})^5-\frac{4v_{0}^3v_{3}}{\gamma_{1}^3}(\xi_{\mathbf{k}}^{\dagger})^3(\xi_{\mathbf{k}})+\frac{3v_{0}^2}{\gamma_{1}^2}(\xi_{\mathbf{k}}^{\dagger})^2(\frac{\gamma_{2}}{2})+\frac{3v_{0}v_{3}^2}{\gamma_{1}^2}(\xi_{\mathbf{k}}^{\dagger})(\xi_{\mathbf{k}})^2-\frac{2v_{3}}{\gamma_{1}}(\xi_{\mathbf{k}}^{\dagger})(\frac{\gamma_{2}}{2}).
\end{equation}
The low-energy bands near the $K$ valley for RPG without and with $\gamma_3$ are shown in Fig.~\ref{fig_g3} (b) (and (c)). 
From these figures, we can conclude that the $\gamma_3$-trigonal warping term induces further splitting of the central touching point. However, while this splitting transforms the quadratic band-touching at the \(K\) point into a central Dirac cone and three satellite Dirac cones located at and around the \(K\) point, the momentum amplitude for the three satellite Dirac cones remains very small due to the condition \(v_0 \gg v_3\). Consequently, the condition for mass inversion in the central Dirac cone, $\Delta = -m$, is very close to the condition for mass inversion in the satellite Dirac cones, $D = \frac{2\epsilon - 1}{1 + \epsilon}m$, where $\epsilon = \frac{v_0^2 a_{0}^2 k^2}{2\gamma_1^2} \ll 1$. This proximity results in the same phase diagram as in the case of $\gamma_3 = 0$ when the displacement field increases at a fixed staggered layer order.

\subsection{I-D. The entire phase diagram under increasing displacement field for RPG}
\begin{figure}[h]
    \centering
    \includegraphics[width=0.8\linewidth]{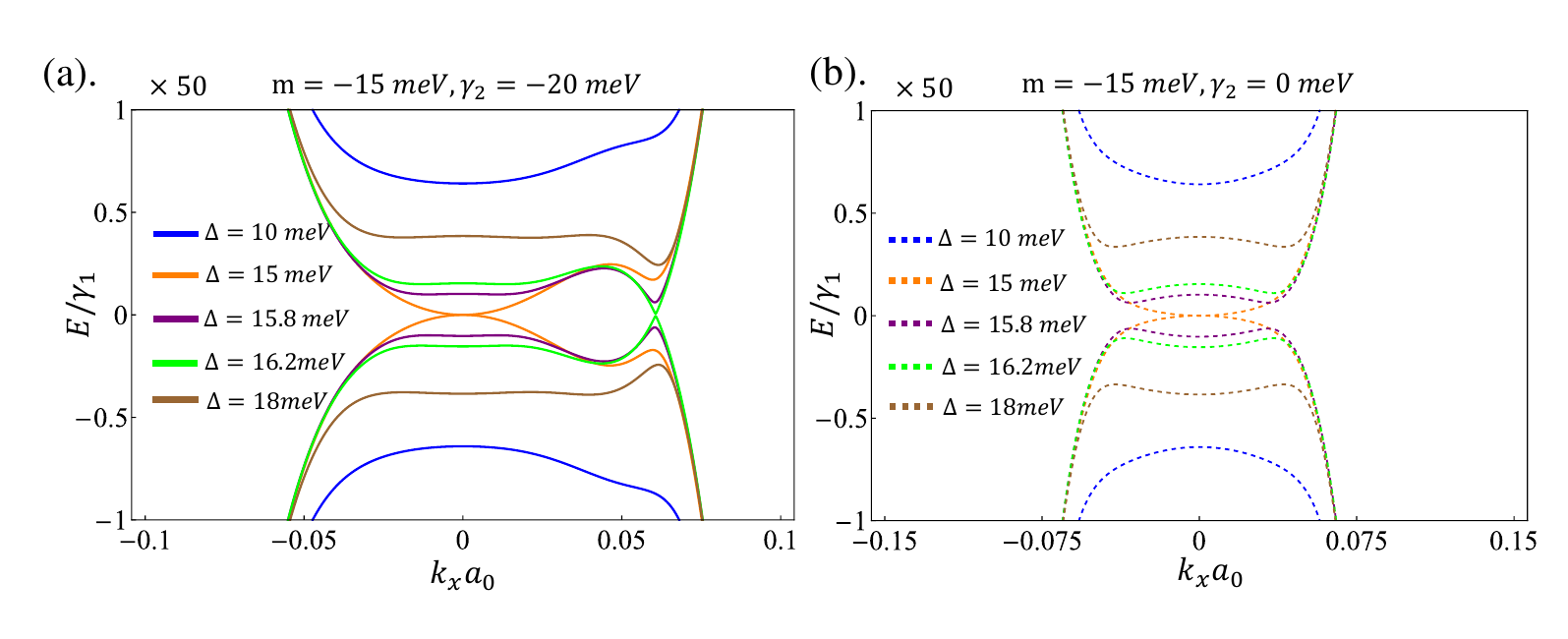}
    \caption{(a). The evolution of the band structure with increasing interlayer potential at $K$ valley, in the presence of the staggered sublattice order given in Eq.~(\ref{ano-inter}). The parameters are $m=m_{1}=-15 ~meV$, $\gamma_{2}=-20 ~meV$, $\gamma_{0}=3160~ meV$, and $\gamma_{1}=390~ meV$. (b). The evolution of the band structure with increasing interlayer potential in the case where $\gamma_{2}=0$. All other parameters are the same as in (a).}
    \label{rpg_evolution2}
\end{figure}
Using the staggered layer order defined in Eq.~(\ref{interorder}), we can explain the topological phase transitions from the \(C = 0\) state to the \(C = -3\) QAH state, and then to the \(C = -5\) QAH state. However, as the displacement field continues to increase, the transition from the \(C = -5\) to the \(C = -3\) QAH state, and then to the \(C = 0\) state, remains unexplained~\cite{han2024correlated}. To account for this sequence of topological phase transitions, it is necessary to introduce an order parameter that first induces mass inversion at the central point. One natural choice is the staggered sublattice order, which describes a uniform charge density across each layer but with opposite charge polarizations between the \(A\) and \(B\) sublattices within a single layer. This order also satisfies the charge neutrality condition. The specific form of the Hamiltonian for this order parameter in the basis $ \psi=(A_1,B_1,A_2,B_2,\dots,A_5,B_5)$ is written as:
\begin{equation}\label{ano-inter}
    H'_{m}=m ~Diag(\sigma_{z},\sigma_{z},...,\sigma_{z}),
\end{equation}
where $\sigma_{z}$ is the Pauli matrix of the sublattice degree of freedom on each layer. By performing perturbation theory, the effective mass for the low-energy effective model is derived as:
\begin{equation}\label{ano-effm}
     M'_{5,\text{eff}}(k)=
    (m+\Delta)+\frac{(2 m+\Delta)v_{0}^2k^2}{ 2\gamma_{1}^2}.
\end{equation}
For $k=0$ point, the effective mass changes sign at $\Delta=\Delta'_{1}=-m$ (assuming $\Delta>0$ and $m<0$). However, for $k \neq 0$ points, unlike the case of the staggered layer order, the effective mass changes sign at $\Delta=\Delta'_2=-m\frac{1+2\epsilon}{1+\epsilon}$ which is larger than $\Delta'_{1}=-m$ due to $\epsilon=\frac{v_{0}^2 a_{0}^2 k^2}{2\gamma_{1}^2}>0$. Thus, as the displacement field increases, the effective mass at the central point of the $K$ (or $K'$) valley changes sign first at $\Delta=\Delta_{1}'=-m$, followed by the sign change of the effective mass of the three satellite Dirac cones at $\Delta=\Delta'_2>\Delta_{1}'$. In the presence of the staggered sublattice order, the evolution of the band structure at the \(K\) valley with increasing interlayer potential is shown in Fig.~\ref{rpg_evolution2} (a) and (b), for cases with and without trigonal warping, respectively. 

We then assume that the staggered orders for the four flavors $K \uparrow$, $K \downarrow$, $K^{'} \uparrow$, and $K^{'} \downarrow$ are given as $H'_{m_{1}}$, $-H'_{m_{1}}$, $H_{m_{2}}$, and $-H_{m_{2}}$, respectively, with $0 > m_{2} > m_{1}$. Under this assumption, the order in the \(K\) valley corresponds to staggered sublattice order, while in the \(K'\) valley, it corresponds to staggered layer order. When the interlayer potential satisfies \(0 < \Delta < -m_{2}\), as \(\Delta\) increases, mass inversions occur at the three satellite Dirac cones for \(K'\uparrow\) flavor, causing the Chern number of RPG to change from \(C = 0\) to \(C = -3\). Then, when \(\Delta = -m_{2}\), the effective mass at the central band-touching point for electrons with flavor $K' \uparrow$ changes sign, leading to a further change in the Chern number from \(C = -3\) to \(C = -5\). When the displacement field satisfies $-m_{2} < \Delta < -m_{1}$, since $m_{2} > m_{1}$, the Chern number for RPG remains $C = -5$. As the displacement field continues to increase, for electrons with flavor $K \uparrow$, the effective mass of the central band-touching point changes sign at $\Delta = -m_{1}$, leading to a topological phase transition from the QAH state with $C = -5$ to $C = -3$. With further increases in the displacement field, the effective masses of the three satellite Dirac cones with flavor $K \uparrow$ change sign, resulting in a topological phase transition from the QAH state with $C = -3$ to the state with $C = 0$. Consequently, we deduce that as the displacement field increases, the Chern number for RPG evolves as $0 \rightarrow -3 \rightarrow -5 \rightarrow -3 \rightarrow 0$, which is consistent with experimental observations~\cite{han2024correlated}. Moreover, within this mechanism, we can derive that at \(\Delta = 0\), the \(C = 0\) state in RPG corresponds to a state where both the spin and valley Chern numbers are 0. When the displacement field becomes sufficiently large, although the total Chern number of RPG remains 0, the system transitions into a quantum valley Hall state with a valley Chern number of 5.

\section*{\bf{\uppercase\expandafter{Supplemental Material NOTE II: Bernal multilayer graphene}}}

\subsection{II-A. low-energy effective model for Bernal multilayer graphene}

In this section, we derive the low-energy effective model for Bernal multilayer graphene (BMG). For Bernal \(N\)-layer graphene, if \(N\) is even, the Hamiltonian can be decomposed into \(N/2\) mutually coupled bilayer graphene systems~\cite{koshino2011landau}. We can then project the Hamiltonian for each bilayer graphene onto the basis which is composed of non-dimerized ``atoms" to obtain an \(N\)-dimensional low-energy effective model. If \(N\) is odd, the low-energy effective model for BMG consists of \(N/2\) mutually coupled bilayer graphene systems along with an additional monolayer graphene that couples to these bilayer graphenes. The presence of this additional monolayer graphene results in a metallic state, even when staggered layer orders are considered~\cite{grushina2015insulating}, thereby preventing the formation of a QAH state. As a result, this paper focuses on even-layer BMGs.
The full Hamiltonian for \(N\)-layer BMG in the basis $\psi=(A_1,B_1,A_2,B_2,\dots,A_N,B_N)$ is written as:
\begin{equation}\label{fullHBMG}
     H=\begin{pmatrix}
       H_0 & V & W & & & & \\
        V^\dagger & H'_0 & V^\dagger & W' & & & \\
          W & V & H_0 & V & W & & \\
             & W' & V^\dagger & H'_0 & V^\dagger & W' & \\
               & & \ddots & \ddots & \ddots & \ddots & \ddots
        \end{pmatrix},
\end{equation}
where the intralayer hopping is expressed by $H_{0}=\begin{pmatrix} 0&v_0 \xi_{\mathbf{k}}^{\dagger}\\v_0\xi_{\mathbf{k}} & \delta \end{pmatrix}$, and $H'_{0}=\begin{pmatrix} \delta &v_0 \xi_{\mathbf{k}}^{\dagger}\\v_0\xi_{\mathbf{k}} & 0 \end{pmatrix}$. In these equations, $\xi_{\mathbf{k}}=\zeta k_x+i k_y$ ($\zeta= \pm 1$ representing the K and K' valley, respectively), and $\delta$ represent the on-site energy for the dimerized atoms. The nearest interlayer hopping is represented by $V=\begin{pmatrix} -v_4 \xi_{\mathbf{k}}^{\dagger}& v_3 \xi_{\mathbf{k}} \\ \gamma_1 & -v_4 \xi_{\mathbf{k}}^{\dagger} \end{pmatrix}$, and the next-nearest interlayer hoppings are written as $W=\begin{pmatrix} \gamma_2/2 & 0 \\ 0 & \gamma_{5}/2 \end{pmatrix}$ and $W'=\begin{pmatrix} \gamma_5/2 & 0 \\ 0 & \gamma_{2}/2 \end{pmatrix}$. The side view and hopping parameters are labeled in Fig.~\ref{BMG_s2}~(a) ($v_{i}=\sqrt{3}\gamma_{i}/2$, in this figure, the tetralayer graphene is shown as an example).
\begin{figure}[h]
    \centering
    \includegraphics[width=0.55\linewidth]{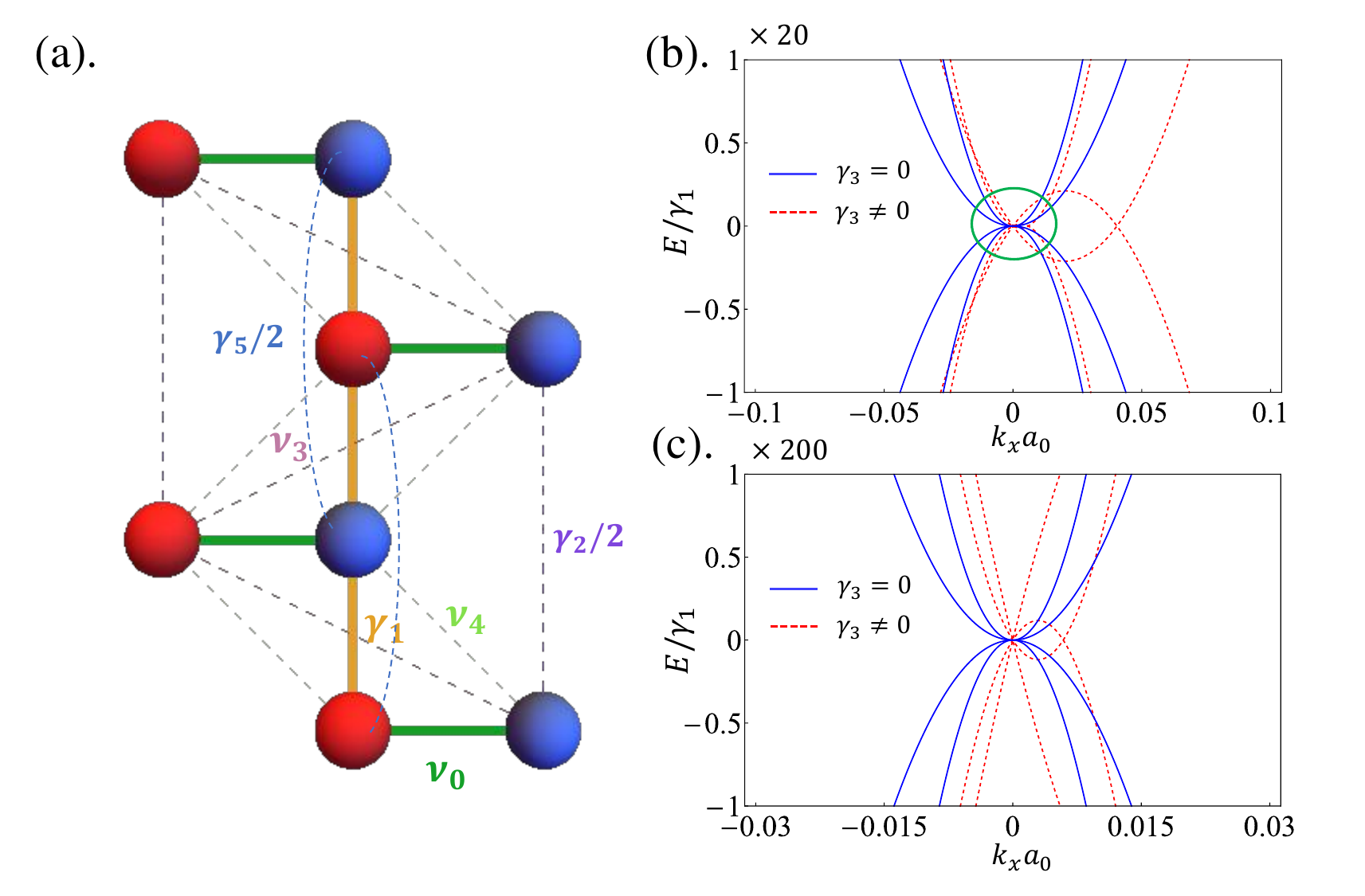}
    \caption{(a). Side view of the Bernal tetralayer graphene (BTG), the blue and red atoms represent the $A$ and $B$ sublattices, respectively. The hopping parameters are indicated by different colors: $\gamma_{0}=3000~ meV$, $\gamma_{1}=400~ meV$, $\gamma_{2}=-20~meV$, $\gamma_{4}=40~meV$, $\gamma_{5}=40~meV$, $\delta=40.8~meV$. Here, the displacement field and staggered layer orders are set to zero: $\Delta=0$, and $m=0$. (b). The band structure for the low-energy effective model with and without $\gamma_{3}=300 ~meV$, which are shown by red dashed and blue solid lines, respectively. (c). Enlarged view of the band structure for the region marked by the green circle in (b).}
    \label{BMG_s2}
\end{figure}

For $N$-layer BMG, as discussed in Ref.\cite{koshino2011landau}, we can decompose the Hamiltonian into a nearly block-diagonal form by using a unitary transformation. For $N \in Even$, the new basis is constructed by combining the old basis as follows:
\begin{equation}
    \phi^{new}=(\phi^{A,odd}_{1},\phi^{B,odd}_{1},\phi^{A,even}_{1},\phi^{B,even}_{1},...,\phi^{A,odd}_{\alpha},\phi^{B,odd}_{\alpha},\phi^{A,even}_{\alpha},\phi^{B,even}_{\alpha},...,\phi^{A,odd}_{N-1},\phi^{B,odd}_{N-1},\phi^{A,even}_{N-1},\phi^{B,even}_{N-1}),
\end{equation}
where $\phi^{X,odd}_{\alpha}=\sum_{j=1}^{N}f_{\alpha}(j)X_{j}$, $\phi^{X,even}_{\alpha}=\sum_{j=1}^{N}g_{\alpha}(j)X_{j}$, and $\alpha=1,3,5,...,N-1$. In these equations, $X=A$ or $B$, $j$ labels the layers, $f_{\alpha}(j)=\sqrt{\frac{1}{N+1}}(1-(-1)^{j})\sin(\kappa_{\alpha}j)$, and $g_{\alpha}(j)=\sqrt{\frac{1}{N+1}}(1+(-1)^{j})\sin(\kappa_{\alpha}j)$, where $\kappa_{\alpha}=\frac{\pi}{2}-\frac{\alpha \pi}{2(N+1)}$. In addition, during the decomposition, we consider the staggered layer order given in Eq.(\ref{interorder}) and the interlayer potential induced by the displacement field, as shown in Eq.(\ref{displacement}). After decomposition, the resulting Hamiltonian can be interpreted as $N/2$ mutually coupled bilayer graphenes. For each bilayer graphene, the low-energy bands can be obtained by projecting the Hamiltonian onto the basis composed of non-dimerized ``atoms":
\begin{equation}
    \phi^{new,low}=(\phi^{A,odd}_{1},\phi^{B,even}_{1},...,\phi^{A,odd}_{\alpha},\phi^{B,even}_{\alpha},...,\phi^{A,odd}_{N-1},\phi^{B,even}_{N-1}).
\end{equation}
Therefore, by using the condition $\gamma_{0},\gamma_{1},\gamma_{3}>> |\gamma_{2}|,\gamma_{4},\gamma_{5},\delta$, the low-energy effective model for $N$-layer BMG can be written as:
\begin{equation}
    H=\begin{pmatrix}
       h_{eff}(\lambda_{1})+h_{\Delta}(s_{1,1},t_{1,1}) & h_{c}(s_{1,3},t_{1,3}) & h_{c}(s_{1,5},t_{1,5}) & ... \\
        h_{c}(s_{1,3},t_{1,3}) & h_{eff}(\lambda_{3})+h_{\Delta}(s_{3,3},t_{3,3}) & h_{c}(s_{3,5},t_{3,5}) & ...  \\
          h_{c}(s_{1,5},t_{1,5}) & h_{c}(s_{3,5},t_{3,5}) & h_{eff}(\lambda_{5})+h_{\Delta}(s_{5,5},t_{5,5}) & ...   \\
            \vdots   & \vdots & \vdots & \ddots &  \\ 
        \end{pmatrix},
\end{equation}
where $h_{eff}(\lambda)=\begin{pmatrix} 0&\lambda v_3 \xi_{\mathbf{k}}-\frac{1}{\lambda \gamma_{1}} v^2_0 (\xi_{\mathbf{k}}^{\dagger})^2 \\ \lambda v_3 \xi_{\mathbf{k}}^{\dagger}-\frac{1}{\lambda \gamma_{1}} v^2_0 (\xi_{\mathbf{k}})^2 & 0 \end{pmatrix}$, $h_{\Delta}(s,t)=\begin{pmatrix} s\Delta+m&0 \\ 0 & t\Delta-m \end{pmatrix}$, and $h_{c}(s,t)=\begin{pmatrix} s\Delta&0 \\ 0 & t\Delta \end{pmatrix}$. In these equations, $\lambda_{\alpha}=2\cos(\kappa_{\alpha})$, $s_{\alpha,\beta}=\sum_{l'}^{N/2}f_{\alpha}(2l'-1)f_{\beta}(2l'-1)(1-\frac{4l'-4}{N-1})$, and $t_{\alpha,\beta}=\sum_{l'}^{N/2}g_{\alpha}(2l')g_{\beta}(2l')(1-\frac{4l'-2}{N-1})$.

For Bernal tetralayer graphene (BTG), the low-energy Hamiltonian can be written as:
\begin{equation}
    H_{B-tetra}=\begin{pmatrix}
       h_{eff}(\lambda_{1})+h_{\Delta}(s_{1,1},t_{1,1}) & h_{c}(s_{1,3},t_{1,3}) \\
        h_{c}(s_{1,3},t_{1,3}) & h_{eff}(\lambda_{3})+h_{\Delta}(s_{3,3},t_{3,3})   
        \end{pmatrix},
\end{equation}
where $\lambda_{1}=(\sqrt{5}-1)/2$, $\lambda_{3}=(\sqrt{5}+1)/2$, $s_{1,1}=-t_{1,1}=\frac{1}{15}(5+2\sqrt{5})$, $s_{3,3}=-t_{3,3}=\frac{1}{15}(5-2\sqrt{5})$, and $s_{1,3}=t_{1,3}=\frac{4}{3\sqrt{5}}$. The band structures for the low-energy effective model with and without \(\gamma_{3} = 300~\text{meV}\) are shown in Fig.~\ref{BMG_s2}~(b) (and (c)). Here, the interlayer potential \(\Delta\) induced by the displacement field and the staggered layer order \(m\) are set to zero.

\begin{figure}[h]
    \centering
    \includegraphics[width=0.65\linewidth]{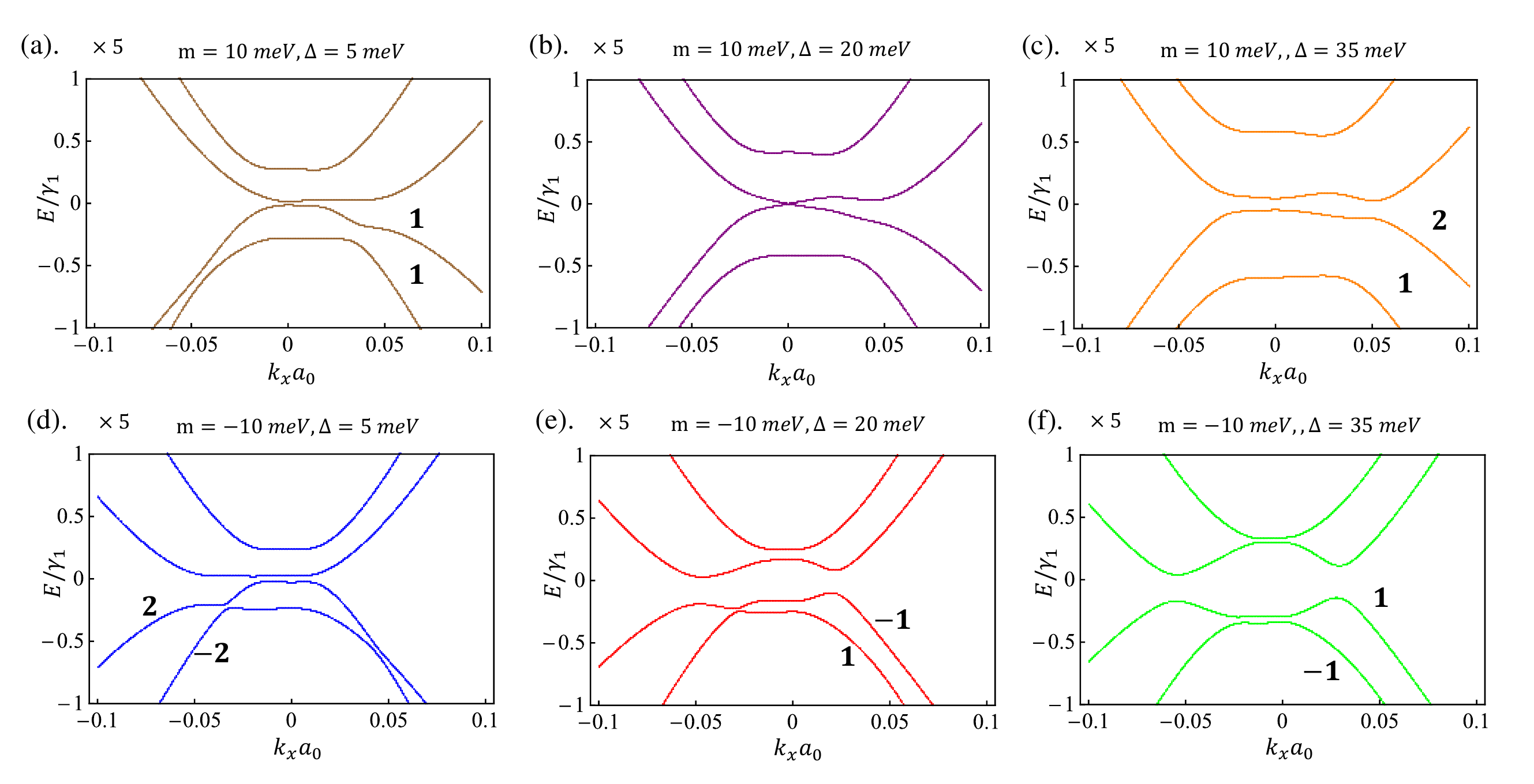}
    \caption{(a), (b), (c). The band structure for the full model of Bernal tetralayer graphene under the increasing displacement field in the presence of the staggered layer orders $m=10~meV$ in $K$ valley. (d), (e), (f). The band structure for the full model of Bernal tetralayer graphene under the increasing displacement field in the presence of the staggered layer orders $m=-10~meV$ in $K'$ valley. The Chern number for each valence band is indicated near the respective band. The hopping parameters are: $v_{0}=\frac{\sqrt{3}}{2} \times 3000~ meV $, $\gamma_{1}=400~meV$, $\gamma_{2}=-20~meV$, $v_{3}=\frac{\sqrt{3}}{2}\times 300~meV$, $v_{4}=\frac{\sqrt{3}}{2}\times 40~meV$, and $\gamma_{5}=40~meV$.}
    \label{figs4}
\end{figure}

\subsection{II-B. The band structure and topological phase transition for the full model of the Bernal tetralayer graphene}

In this section, we calculate the band structure for the full model of Bernal tetralayer graphene (Eq.~(\ref{fullHBMG}) with the conduction \(N = 4\)), and investigate the topological phase transition that occurs as the displacement field increases in the presence of staggered layer orders (Eq.~(\ref{interorder})). From the results shown in Fig.~\ref{figs4}, while the band structure and topological phase transition points differ from those in the low-energy effective model, the QAH state with a total Chern number of \(C = 6\) persists.

\begin{figure}[h]
    \centering
    \includegraphics[width=0.65\linewidth]{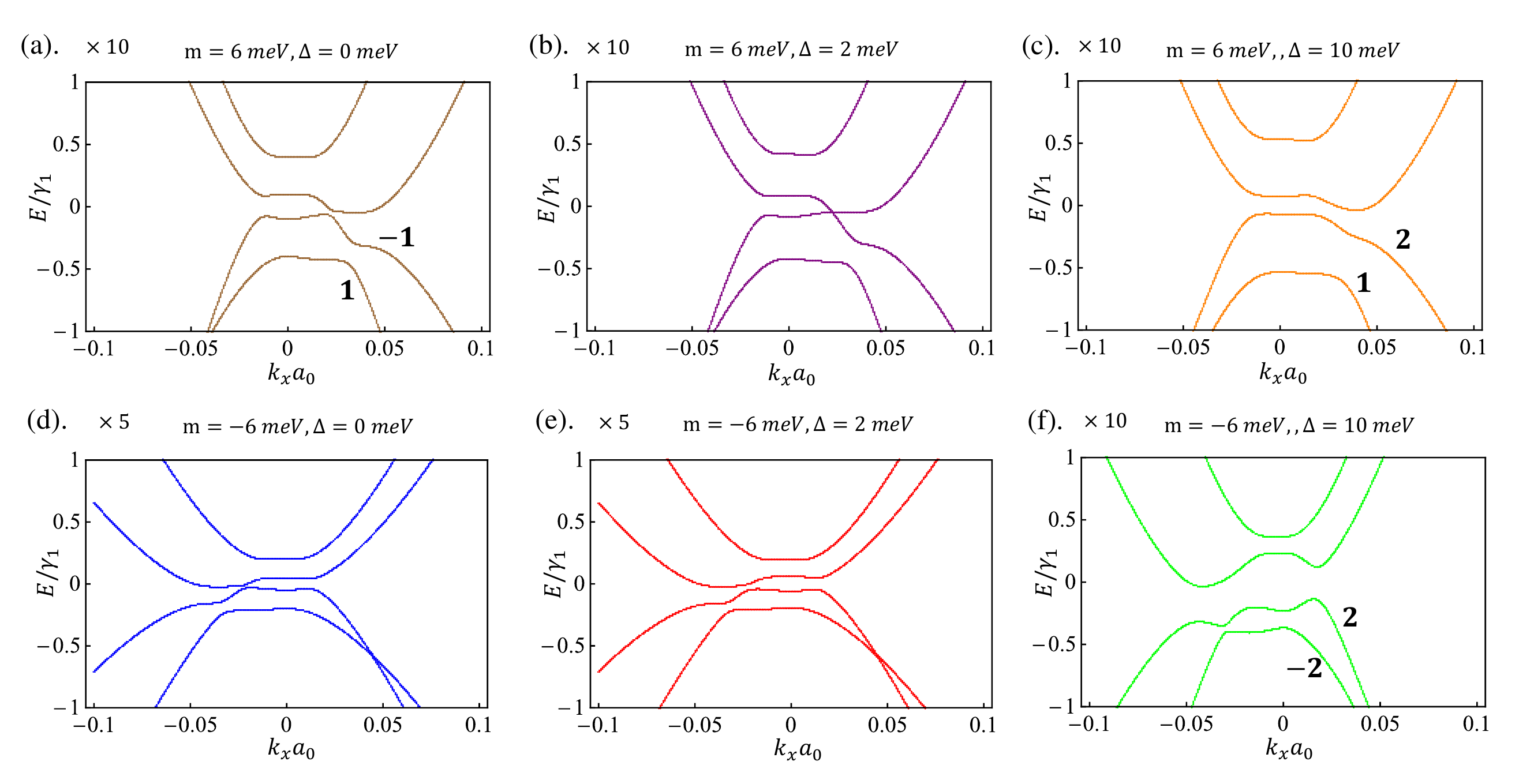}
    \caption{(a), (b), (c). The band structure for the full model of Bernal tetralayer graphene under the increasing displacement field in the presence of the staggered layer orders $m=6~meV$ in $K$ valley. (d), (e), (f). The band structure for the full model of Bernal tetralayer graphene under the increasing displacement field in the presence of the staggered layer orders $m=-6~meV$ in $K'$ valley. The Chern number for each valence band is indicated near the respective band. The hopping parameters are: $v_{0}=\frac{\sqrt{3}}{2} \times 3000~ meV $, $\gamma_{1}=400~meV$, $\gamma_{2}=-20~meV$, $v_{3}=\frac{\sqrt{3}}{2}\times 300~meV$, $v_{4}=\frac{\sqrt{3}}{2}\times 40~meV$, and $\gamma_{5}=40~meV$.}
    \label{figs5}
\end{figure}

Additionally, the topological phases and band structure in the full model of BTG are influenced by the amplitude of the staggered layer order. In Fig.~\ref{figs5}, compared to the case in Fig.~\ref{figs4}, we choose staggered layer orders for the \(K\) and \(K'\) valleys with smaller amplitudes (\(|m| = 6~ meV\)). It turns out that the QAH state with Chern number \(C = 4\) disappears, while the QAH state with Chern number \(C = 6\) persists, which is consistent with the experimental observations~\cite{cho2025tunable}. In the case of \(|m| = 6~meV\), the next-nearest interlayer hoppings \(\gamma_2\) and \(\gamma_5\) significantly affect the effective masses of the first and second bilayer graphene, preventing the mass inversion at the central Dirac cone of the first bilayer graphene. The topological phase transition in the \(K\) valley is driven by the mass inversion of the satellite Dirac cones belonging to the bands of the first bilayer graphene,  resulting in a change of the Chern number by \(3\).

\end{widetext}
\end{document}